\documentclass[12pt]{article}

\usepackage{epsf}

\begin{document}

\begin{center}

{\large\bf Tensor analyzing power $A_{yy}$ in deuteron inclusive breakup at
large $P_t$ and spin structure of deuteron at short internucleonic
distances.\footnote{\it Talk given at the SPIN2004 Conference, 
10-16 October 2004, Triest, Italy}$^,$
\footnote{\uppercase{T}his work is supported by the \uppercase{RFFR} 
under grant \uppercase{N}o. 03-02-16224}}
\vskip 5mm

{V.P.Ladygin\footnote{\uppercase{E}-mail address: ladygin@sunhe.jinr.ru},
L.S.~Azhgirey, S.V.~Afanasiev, V.V.~Arkhipov, V.K.~Bondarev,
Yu.T.~Borzounov, 
L.B.~Golovanov, A.Yu.~Isupov, A.A.~Kartamyshev,
V.A.~Kashirin, A.N.~Khrenov, V.I.~Kolesnikov,
V.A.~Kuznezov, A.G.~Litvinenko,
S.G.~Reznikov, P.A.~Rukoyatkin, A.Yu.~Semenov, 
I.A.~Semenova, G.D.~Stoletov, A.P.~Tzvinev,
 V.N.~Zhmyrov and  L.S.~Zolin}

{\it Joint Institute for Nuclear Researches, 141980 Dubna, Russia}
\vskip 3mm

{G.Filipov} \\

{\it Institute of Nuclear Research and Nuclear Energy, 1784 Sofia, 
Bulgaria}
\vskip 3mm

{N.P.~Yudin}\\

{\it Moscow State University, 117234 Moscow, Russia}  

\end{center}

\begin{abstract}{
The $A_{yy}$ data for deuteron inclusive breakup off hydrogen
and carbon at 
a deuteron momentum  
of 9.0 GeV/c and
large $p_T$ 
of emitted protons are presented.
The large values of $A_{yy}$ independent 
of the target mass reflect the sensitivity of the data 
to the deuteron spin structure. 
The data obtained at fixed   
$x$ and plotted versus $P_t$ clearly
demonstrate the dependence of the deuteron spin structure 
at short internucleonic distances on  
two variables. 
The data are compared with the calculations using Paris, CD-Bonn and 
Karmanov's deuteron wave functions.}
\end{abstract}

\section{Introduction}

The interest to the $(d,p)$ reaction at relativistic energies 
is  
mostly due to the
possibility to observe
the manifestation of the non-nucleonic degrees of freedom and 
relativistic effects in the simplest bounded system.

Large amount of the polarization data in deuteron breakup obtained 
at a zero degree
last years can be interpreted from the point of view 
$NN^*$ configurations  in the deuteron, where relativistic effects
are taken into account by the minimal relativization scheme with
the dependence of the deuteron structure on single variable $k$.
In addition the considering of  multiple scattering is  required
to obtain the agreement with the data\cite{kob}.

On the other hand,
it was shown that $T_{20}$ data for the pion-free 
deuteron breakup process $dp \rightarrow ppn$ in the kinematical 
region close to that of backward elastic $dp$ scattering depended 
on the incident deuteron momentum in addition to $k$\cite{azh1}. 
The recent measurements of the tensor analyzing 
power $A_{yy}$ of deuteron inclusive breakup  on nuclear 
targets \cite{ayy90,ayy45,ayy50}
have demonstrated a significant dependence on the transverse 
secondary proton momentum  $p_{T}$ being plotted at a fixed 
value of the longitudinal proton momentum.
This forces 
one to suggest that description of this quantity requires an 
additional independent variable, aside from $k$.  

In this report the angular dependence of $A_{yy}$ in deuteron inclusive 
breakup on hydrogen and carbon at 9 GeV/c are presented.
The results are compared with the relativistic calculations using Paris, 
CD-Bonn and 
Karmanov's deuteron wave functions (DWFs).

\section{Experiment}\label{sec:exp}

The experiment has been performed using a tensorially polarized
deuteron beam of the Dubna Synchrophasotron and the SPHERE setup described
elsewhere \cite{ayy90,ayy45}.
 The tensor polarization of the beam has been 
measured from the asymmetry of protons ~from the~ deuteron~
breakup ~on nuclear~ targets, $d+A \to p+X$, at a 
zero angle and the momenta $p_p\sim 2/3p_d$ \cite{zolin}.
The vector polarization of the beam has been measured from the asymmetry
of quasi-elastic $pp$ scattering on $CH_2$ target \cite{f4}.
The tensor and vector polarizations, $p_{zz}$ and $p_{z}$, were
$p_{zz}^+=0.798\pm 0.002(stat)\pm 0.040(sys)$,
$p_{zz}^-=-0.803\pm 0.002(stat)\pm 0.040(sys)$,
$p_z^+=0.275\pm 0.016(stat)\pm 0.014(sys)$ and
$p_z^-=0.287\pm 0.016(stat)\pm 0.014(sys)$, respectively.

A slowly extracted deuteron beam with a typical intensity 
of $\sim 5\cdot 10^8\div 10^9$ $\vec{d}$/spill was directed onto
a liquid hydrogen target of 30 cm  
length or onto carbon targets
with varied length.  The data at 9 GeV/c 
of the  deuteron
initial momentum were obtained
at secondary proton emission angles of 85, 130 and 160 mr and
proton momenta between 4.5 and 7.0 GeV/c on hydrogen and carbon.
The separation
of the protons and inelastically scattered deuterons  was done by 
the measurements of their time-of-flight (TOF) over a base line of 
$\sim 34$ m. The residual background was completely eliminated by 
the requirement that particles are detected at least in two prompt 
TOF windows.

\section{Results and discussion}

\begin{figure}[ht]
\centerline{\epsfxsize=2.5in\epsfbox{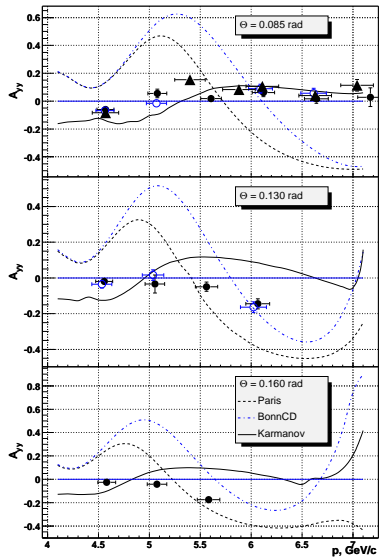}} 
\vskip -7mm  
\caption{ The $A_{yy}$ data plotted versus secondary proton momentum.
The curves are the relativistic calculations with different DWFs. \label{fig1}}
\end{figure}

The results on $A_{yy}$ versus the momentum of the secondary protons 
are presented in 
Fig.~\ref{fig1} by the solid and open symbols 
for carbon and hydrogen targets, respectively. 
The circles are the data of this experiment, while the triangles represent
the data obtained earlier\cite{ayy90}.
The dashed, dash-dotted and solid lines are the relativistic calculations
using Paris\cite{paris}, CD-Bonn\cite{cd-bonn} and Karmanov\cite{karm} DWFs. 

\vskip -5mm

\vskip -5mm

The data obtained on hydrogen and carbon are in good agreement.
Hence  multiple 
scattering processes play a minor role and the obtained 
information reflects the internal deuteron structure.

The calculations performed 
in the framework of the light-front dynamics \cite{azh_yud}
with the use of 
Paris and CD-Bonn DWFs fail to reproduce the $A_{yy}$ data, while
the use of Karmanov DWF depending on  
two internal variables, $k$ and 
$p_T$, is in a reasonable agreement
with the data obtained at 85 mr. 

$A_{yy}$ data obtained at fixed proton momenta of 6.0, 6.5 and 7.0 ~GeV/c
are plotted versus transverse
proton momenta  
$p_T$. The symbols and  curves are the same as in Fig.\ref{fig1}.
Again the use of DWF depending on 
two variables\cite{karm} gives better
agreement with the data. 

To summarize, the observed features of the $A_{yy}$ data suggest that 
the deuteron structure function at short distances, where relativistic
effects are significant,  depends on more than one 
variable.


\vskip -10mm

\begin{figure}[ht]
\centerline{\epsfxsize=2.3in\epsfbox{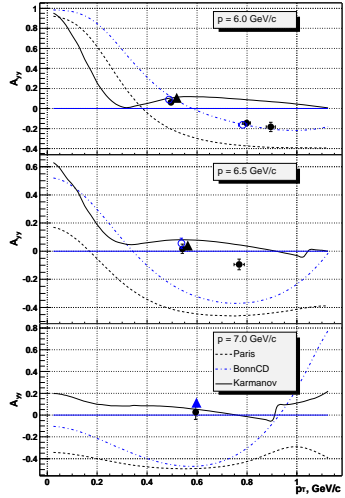}}
\vskip -7mm   
\caption{ The $A_{yy}$ data  obtained at fixed proton momenta in lab. and 
plotted versus transverse proton momentum  
$p_T$.}
\end{figure}

\end{document}